\documentstyle[aps,preprint,prl,epsfig]{revtex}
\newcommand{\beq}{\begin{equation}}
\newcommand{\eeq}{\end{equation}}  
\begin{document}
\tightenlines
\draft

\title{Non Asymptotic Properties of Transport and Mixing}

\author{G. Boffetta$^1$, A. Celani$^1$, M. Cencini$^2$, G. Lacorata$^3$
and A. Vulpiani$^2$}

\address{$^1$ Dipartimento di Fisica Generale and 
Istituto Nazionale Fisica della Materia,\\
Universit\`a di Torino, Via Pietro Giuria 1, 10125 Torino, Italy }
\address{$^2$Dipartimento di Fisica and Istituto 
Nazionale Fisica della Materia\\
Universit\`a di  Roma ``la Sapienza'', 
Piazzale Aldo Moro 5, 00185 Roma, Italy }
\address{$^3$Dipartimento di Fisica, Universit\`a dell' Aquila, \\
Via Vetoio 1, 67010 Coppito, L'Aquila, Italy }

\date{\today}

\maketitle

\begin{abstract}
We study relative dispersion of passive scalar in non-ideal cases,
i.e. in situations in which asymptotic techniques cannot
be applied; typically when the characteristic length scale of the Eulerian
velocity field is not much smaller than the domain size.
Of course, in such a situation usual asymptotic quantities (the
diffusion coefficients) do not give any relevant information about the
transport mechanisms. On the other hand, we shall show that the Finite
Size Lyapunov Exponent, originally introduced for the predictability
problem, appears to be rather powerful in approaching the non-asymptotic
transport properties. This technique is applied in a series of
numerical experiments in simple flows with chaotic behaviors, in
experimental data analysis of drifter and to study relative dispersion
in fully developed turbulence.
\end{abstract}

\newpage

{\bf 
It is now well known that the Lagrangian motion of test particles
in a fluid can be highly non trivial  even for simple Eulerian field. 
Up to now there exist powerful methods to study in a rigorous way the
asymptotic transport properties of passive scalar. On the other hand
very often (especially in real world) it is not possible to characterize
dispersion in terms of asymptotic quantities such as average
velocity and diffusion coefficients. This happens, typically, in 
finite domain systems with no large scale-separation between
the domain size and the largest characteristic Eulerian length;
more generally when there is not a sharp separation among the
characteristic length scales of the system. In this perspective, we
briefly review  a recently introduced method to approach
the non-asymptotic properties of transport and mixing. We discuss the
relevance of the Finite Size Lyapunov Exponent for the
characterization of diffusion. In particular we stress its advantages
compared with the usual way of looking at the relative dispersion at
fixed delay time.
}

\section{Introduction}
\label{sec:1}

Transport processes play a crucial role in many natural phenomena.
Among the many examples, we just mention the particle transport in
geophysical flows which is of obvious interest for atmospheric and
oceanic issues. The most natural framework for investigating such
phenomena is to adopt a Lagrangian viewpoint in which the particles
are advected by a given Eulerian velocity field $\bbox{u}(\bbox{x},t)$
according to the differential equation
\begin{equation}
{d \bbox{x} \over d t} = \bbox{u}(\bbox{x},t) = \bbox{v}(t)\,,
\label{eq:1.1}
\end{equation}
where, by definition, $\bbox{v}(t)$ is the Lagrangian particle velocity.

Despite the apparent simplicity of (\ref{eq:1.1}), the problem of
connecting the Eulerian properties of $\bbox{u}$ to the 
Lagrangian properties of the trajectories $\bbox{x}(t)$
is a very difficult task. In the last $20-30$ years the
scenario has become even more complex by the recognition of
the ubiquity of Lagrangian chaos (chaotic advection). 
Even very simple Eulerian fields can generate very complex
Lagrangian trajectories which are practically indistinguishable from 
those obtained in a complex, turbulent, flow \cite{H66,Licht,Ottino,lagran,zav,CM93}.

Despite these difficulties, the study of the relative 
dispersion of two particles can give some insight on the link
between Eulerian and Lagrangian properties at different length-scales.
Indeed, the evolution of the separation 
$\bbox{R}(t)=\bbox{x}^{(2)}(t)-\bbox{x}^{(1)}(t)$ between two tracers
is given by
\begin{equation}
{d \bbox{R} \over d t} = \bbox{v}^{(2)}(t)-\bbox{v}^{(1)}(t)=
\bbox{u}(\bbox{x}^{(1)}(t)+\bbox{R}(t),t)-
\bbox{u}(\bbox{x}^{(1)}(t),t)
\label{eq:1.5}
\end{equation}
and thus depends on the velocity difference on scale $\bbox{R}$.  It
is obvious from (\ref{eq:1.5}) that Eulerian velocity components of
typical scale much larger than $\bbox{R}$ will not contribute to the
evolution of $\bbox{R}$. Since, in incompressible flows, separation
$\bbox{R}$ typically grows in time \cite{Cocke69,Orszag70} we have the
nice situation in which from the evolution of the relative separation
we can, in principle, extract the contributions of all the components
of the velocity field.  For these reasons, in this paper we prefer to
study relative dispersion instead of absolute dispersion.  For
spatially infinite cases, without mean drift there is no difference;
for closed basins the relative dispersion is, for many aspects, more
interesting than the absolute one, which is dominated by the sweeping
induced by large scale flow.

There are very few general results on the link between Eulerian and
Lagrangian properties and only for asymptotic behaviors.
Let us suppose that the Eulerian velocity field 
is characterized by two typical length-scales:
the (small) scale $l_u$ below which the velocity is smooth, and
a (large) scale $L_0$ representing the size of the largest 
structures present in the flow.
Of course, in most non turbulent flows it will turn out that
$\l_u \sim L_0$.

At very small separations $R \ll \l_u$ we have that the velocity difference
in (\ref{eq:1.5}) can be reasonably approximated by a linear expansion in $R$,
which in most time-dependent flows leads to an exponential growth
of the separation of initially close particles, a
phenomenon known as Lagrangian chaos 
\begin{equation}
\langle \ln R(t) \rangle \simeq \ln R(0) + \lambda t
\label{eq:1.6}
\end{equation}
(the average is taken over many couples with  initial
separation $R(0)$). The coefficient $\lambda$ is the 
Lagrangian Lyapunov exponent of the system \cite{Licht}. 
The rigorous definition of the Lyapunov exponent imposes to take the
two limits $R(0) \to 0$ and then $t \to \infty$: in physical terms
these limits amount to the requirement that the separation
has not to exceed the scale $\l_u$ but for very large times.  
This is a very strict condition, rarely
accomplished in real flows,
rendering often infeasible the experimental observation
of the behavior (\ref{eq:1.6}).

On the opposite limit, for very long times and for separations
$R \gg L_0$, the two trajectories $\bbox{x}^{(1)}(t)$ and $\bbox{x}^{(2)}(t)$
feel two velocities which can practically be considered as uncorrelated. 
We thus expect normal diffusion, i.e.
\begin{equation}
\langle R^2(t) \rangle \simeq 2 D\, t\,.
\label{eq:1.7}
\end{equation}
Also in this case it is necessary to remark that the asymptotic 
behavior (\ref{eq:1.7})
cannot be attained 
in many realistic situations, the most common of which
is the presence of boundaries at a scale comparable with $L_0$.
In absence of boundaries it is possible to formulate 
sufficient conditions on the nature of the Eulerian flow, under which
normal diffusion (\ref{eq:1.7}) always takes place asymptotically 
\cite{MA89-AV95}.

Between the two asymptotic regimes (\ref{eq:1.6}) and (\ref{eq:1.7})
the behavior of $R(t)$ depends on the particular flow.  The study
of the evolution of the relative dispersion in this crossover regime
is very interesting and can give an insight on the Eulerian structure
of the velocity field.

To summarize, in all systems in which the characteristic length-scales 
are not sharply separated, it is not possible to describe dispersion
in terms of asymptotic quantities. In such cases, 
different approaches are required. Let us mention some examples:
the symbolic dynamics approach to the sub-diffusive
behavior in a stochastic layer \cite{A} and to mixing in 
meandering jets \cite{B}; the study of tracer dynamics in open flows in terms
of chaotic scattering \cite{C} and the exit time description for transport
in semi-enclosed basins \cite{D} and open flows \cite{E}.

The aim of the present paper is to discuss the use of an indicator --
the Finite Size Lyapunov Exponents (FSLE), originally introduced in
the context of predictability problems \cite{ABCPV97} -- to study and
characterize non-asymptotic transport in non-ideal systems, e.g.
closed basins and systems in which the characteristic length-scales
are not sharply separated.

In section \ref{sec:2} we introduce the basic tools for the 
finite-scale analysis and we discuss their general properties.
Section \ref{sec:3} is devoted to the evaluation of our method
on some numerical examples. We shall see that even in 
apparently simple situations the use of finite scale analysis 
avoids possible misinterpretation of the results.
In section \ref{sec:4} the method is applied to two physical problems:
the analysis of experimental drifter data and the numerical study of 
relative dispersion in fully developed turbulence. 
Conclusions are presented in section \ref{sec:5}.
The appendices  report, for sake of self-consistency, some technical aspects.

\section{Finite size diffusion coefficient}
\label{sec:2}

In order to introduce the finite size analysis for the dispersion problem
let us start with a simple example. 
We consider a set of N particle pairs advected 
by a smooth (e.g. spatially periodic) velocity field with characteristic
length $l_{u}$. 
Denoting with $R_i^{2}(t)$ the square separation of the $i$-th couple,
we define
\begin{equation}
\langle R^{2}(t) \rangle= 
{1 \over N} \sum_{i=0}^N R_i^2 \, .
\label{def:disprel}
\end{equation}
We assume that the Lagrangian motion is chaotic,
thus we expect the following regimes to hold
\begin{equation}
\langle R^{2}(t) \rangle \simeq \left\{ 
\begin{array}{ll}
R^{2}_{0}\exp(L(2)t) & \;\;\;\;
{\mbox {if    $\langle R^{2}(t)\rangle^{1/2} \ll l_{u}$}}
 \\
2 D t & \;\;\;\;
{\mbox {if    $\langle R^{2}(t)\rangle^{1/2} \gg l_{u}$}}
\end{array}
\label{eq:regimiperR}
\right.
\,,
\label{example1} 
\end{equation}
where $L(2) \geq 2\lambda$ is the generalized Lyapunov exponent
\cite{BPPV85,PV87}, $D$ is the diffusion coefficient and 
we assume that $R_i(0)=R_0$.

An alternative method to characterize the dispersion properties is by
introducing the ``doubling time\/'' $\tau(\delta)$ at scale $\delta$
as follows \cite{ABCCV97}: given a series of thresholds $\delta^{(n)}=
r^{n} \delta^{(0)}$, one can measure the time $T_i(\delta^{(0)})$ it
takes for the separation $R_i(t)$ to grow from $\delta^{(0)}$ to
$\delta^{(1)}= r \delta^{(0)}$, and so on for
$T_i(\delta^{(1)})\,,\;T_i(\delta^{(2)})\,,\ldots$ up to the largest
considered scale.  The $r$ factor may be any value $> \, 1$,
properly chosen in order to have a good separation between the scales
of motion, i.e. $r$ should be not too large. Strictly speaking,
$\tau(\delta)$ is exactly the doubling time if $r\,=\,2$.

Performing the doubling time experiments over the $N$ particle pairs,
one defines the 
average doubling time $\tau(\delta)$ at scale 
$\delta$ as
\begin{equation}
\tau(\delta) = < T(\delta) >_e =\frac{1}{N}
 \sum_{i=1}^{N} T_{i}(\delta)\,.
\label{def:taudelta}
\end{equation}
It is worth to note that the average 
(\ref{def:taudelta}) is different from the usual
time average (see Appendix \ref{app:1}).

Now we can define the Finite Size
Lagrangian Lyapunov Exponent (see \cite{ABCPV97} for a detailed
discussion) in terms of the average doubling time as 
\begin{equation} 
\lambda(\delta)=\frac{\ln r}{\tau(\delta)}\,, 
\end{equation}
which quantifies the average rate of separation between two particles
at a distance $\delta$. Let us remark that $\lambda(\delta)$ is
independent of $r$, for $r$ close to $1$.  For very small separations
(i.e. $\delta \ll l_u$) one recovers the Lagrangian Lyapunov exponent
$\lambda$, 
\begin{equation} \lambda=\lim_{\delta \rightarrow 0}
\frac{1}{\tau(\delta)} \ln r\,.
\label{def:liapfromtau}
\end{equation}
In this framework the finite size diffusion coefficient
\cite{ABCCV97}, $D(\delta)$, dimensionally turns out to be
\begin{equation}
D(\delta)=\delta^{2}\lambda(\delta)\,.
\label{def:fsd}
\end{equation}
Note the absence of the factor $2$, as one may expect from
(\ref{eq:regimiperR}), in the denominator of $D(\delta)$; 
this is because $\tau(\delta)$ is a difference of
times.  For a standard diffusion process $D(\delta)$ approaches the
diffusion coefficient $D$ (see eq. (\ref{eq:regimiperR})) in the limit
of very large separations ($\delta \gg l_u$). This result stems from
the scaling of the doubling times $\tau(\delta) \sim \delta^2$ for
normal diffusion.

Thus the finite size Lagrangian Lyapunov exponent $\lambda(\delta)$ 
behaves as follows: 
\begin{equation}
\lambda(\delta) \sim \left\{ 
\begin{array}{ll}
\lambda & \;\;\;\;
{\mbox {if    $\delta \ll l_{u}$}}
 \\
D/\delta^{2} & \;\;\;\;
{\mbox {if    $\delta \gg l_{u}$}}
\end{array}
\right.
\,,
\label{eq:regimipertau} 
\end{equation}
One could naively conclude, matching the behaviors 
at $\delta \sim l_{u}$, that $D \sim \lambda l_{u}^{2}$.
This is not always true, since one can have a rather large range
for the crossover due to  
nontrivial correlations which can be present in 
the Lagrangian dynamics \cite{lagran}.

One might wonder that the introduction of $\tau(\delta)$ is just
another way to look at $\langle R^{2}(t)\rangle$.  This is true only
in limiting cases, when the different characteristic lengths are well
separated and intermittency is weak.  A similar idea of using times
for the computation of the factor diffusion coefficient in nontrivial
cases was developed in Ref. \cite{previouswork1,previouswork2,sabot}.

If one wants to identify the physical mechanisms acting on a given spatial 
scale, the use of scale dependent quantities is more appropriate 
than time dependent ones. 

For instance, in presence of strong intermittency (which is indeed a rather
usual situation) $R^{2}(t)$ as a function of $t$ can be very different
in each realization.  Typically one has (see figure~1a), 
different exponential growth rates for different
realizations, producing a rather odd behavior of the average
$\langle R^{2}(t)\rangle$ not due to any physical mechanisms. For instance in
figure~1b we show the average $\langle R^{2}(t)\rangle$ versus
time $t$; at large times one recovers the diffusive behavior but at
intermediate times appears an ``anomalous'' diffusive regime which is only due
to the superposition of exponential and diffusive contributions by
different samples at the same time.  On the other hand, by exploiting
the tool of doubling times one has an unambiguous result (see figure~1c) 
\cite{ABCCV97}.

An important physical problem where the behavior of $\tau(\delta)$ is
essentially well understood is the relative dispersion in 
3D fully developed turbulence.  
Here the smallest Eulerian scale $l_u$ is the Kolmogorov scale 
at which the flow becomes smooth. In the inertial range
$l_u < R < L_0$ we expect the {\it Richardson law} to hold
$\langle R^2(t) \rangle \sim t^3$; for separations larger
than the integral scale $L_0$ we have normal diffusion.
In terms of the finite size Lyapunov exponent we thus expect
three different regimes:
\begin{enumerate}
\item 
$\lambda(\delta)=\lambda$ \,\,\,\, for $\delta \ll l_u$
\item
$\lambda(\delta) \sim \delta^{-2/3}$ \,\,\,\, for $l_u \ll \delta \ll L_0$
\item
$\lambda(\delta) \sim \delta^{-2}$ \,\,\,\, for $\delta \gg L_0$
\end{enumerate}

We will see in section \ref{sec:4} than even for large Reynolds
numbers, the characteristic lengths $l_u$ and $L_0$ are
not sufficiently separated and the different scaling regimes
for $\langle R^2(t) \rangle$ cannot be well detected.
The fixed scale analysis in terms of $\lambda(\delta)$ for 
fully developed turbulence presents clear advantages with respect to
the fixed time approach.

\section{Numerical Results on simple flows}
\label{sec:3}
In this section we shall discuss some examples of 
 applications of the above introduced  indicator 
$\lambda(\delta)$ (or equivalently $D(\delta)$) for
simple flows.
The technical and numerical details of the 
finite size Lyapunov  exponent computation are
settled out in Appendix \ref{app:1}.  

In a generic case in addition to the two asymptotic regimes
(\ref{eq:regimipertau}) discussed in section \ref{sec:2}, 
we expect another universal regime due to the
presence of the boundary of given size $L_B$.  For separations close
to the saturation value $\delta_{max} \simeq L_B$ we expect the
following behavior to hold for a broad class of systems \cite{ABCCV97}:
\begin{equation}
 \lambda(\delta)=\frac{D(\delta)}{\delta^{2}} \propto
\frac{(\delta_{max}-\delta)}{\delta} \,.
\label{eq:nearbound}
\end{equation}
The proportionality constant
is given by  the second eigenvalue of the 
Perron-Frobenius operator which is related to the typical time 
of exponential relaxation of tracers' density to uniform distribution
(see Appendix \ref{app:2}).

\subsection{A model for transport in Rayleigh-B\'enard convection}
\label{sec:3.1}
The advection in two dimensional incompressible flows in absence of
molecular diffusion is given by Hamiltonian equation of motion where
the stream function, $\psi$, plays the role of the Hamiltonian:
\begin{equation}
\frac{dx}{dt}=\frac{\partial \psi}{\partial y}\,, \;\;\;
\frac{dy}{dt}=-\frac{\partial \psi}{\partial x}\,.
\label{eq:hamilton}
\end{equation}

If $\psi$ is time-dependent one typically has chaotic advection.
As an example let us consider the time-periodic Rayleigh-B\'enard convection, 
which can be described by the following stream function \cite{gollub}: 
\begin{equation}
\psi(x,y,t)=\frac{A}{k} \sin\left\{ k \left[ x+B \sin(\omega
t)\right]\right\} W(y)\,,
\label{eq:gollubinf}
\end{equation}
where $W(y)$  satisfies rigid 
boundary conditions on the surfaces $y=0$ and $y=a$ 
(we use $W(y)=\sin(\pi y/a)$).
The two surfaces $y=a$ and $y=0$ are the top and bottom
surfaces of the convection cell.
The time dependent term $B\sin(\omega t)$ represents 
lateral oscillations of the roll pattern  
which mimic the even oscillatory instability \cite{gollub}.

Concerning the analysis in terms of the finite size Lyapunov exponent
one has that, if $\delta$ is much smaller than
the domain size, $\lambda(\delta)=\lambda$.  At larger
values of $\delta$ we find standard diffusion
$\lambda(\delta)=D/\delta^{2}$ with good quantitative agreement with
the value of the diffusion coefficient evaluated by the standard
technique, i.e.  using $\langle R^{2}(t)\rangle$ as a function of time
$t$.

In order to study the effects of finite boundaries on the diffusion
properties we confine the tracers' motion in a closed
domain. This can be achieved by slightly
modifying the stream function (\ref{eq:gollubinf}).  We have modulated
the oscillating term in such a way that for $|x|=L_B$ the amplitude of
the oscillation is zero, i.e. $B \rightarrow B \sin(\pi x/L_B)$ with
$L_B=2\,\pi n/k$ ($n$ is the number of convective cells).  In this way
the motion is confined in $x \in [-L_B,L_B]$.

In figure~2 we show $\lambda(\delta)$ for two values of $L_B$. 
If $L_B$ is large enough one can distinguish the three regimes:
exponential, diffusive and the saturation regime 
eq. (\ref{eq:nearbound}).
Decreasing the size of the boundary $L_B$, the range
of the diffusive regime decreases, while for small values of
$L_B$, it disappears. 

\subsection{Point vortices in a Disk}
\label{sec:3.2}
We now consider a two-dimensional time-dependent flow generated by 
$M$ point vortices, with circulations
$\Gamma_1,\dots,\Gamma_M$, in a disk of unit radius \cite{Aref}.
The passive tracers are advected by the time dependent velocity 
field generated by the vortices and behave chaotically for 
any $M>2$.
Let us note that in this case the scale separation is not imposed 
by hand, but depends on $M$ and on the energy of the vortex 
system \cite{BCF96}.
Figure 3a
shows the relative diffusion as a function of time in a system with 
$M=4$ vortices.  
Apparently there is an intermediate regime of anomalous
diffusion.  However from figure 3b one can clearly see that,
with the fixed scale analysis, only two regimes survive: exponential
and saturation.  Comparing figure 3a and figure 3b one
understands that the appearance of the spurious anomalous diffusion
regime in the fixed time analysis is due to the mechanism 
described in section \ref{sec:2}.

The absence of the diffusive regime $\lambda(\delta) \sim \delta^{-2}$
is due to the fact that the characteristic length of the velocity
field, which is comparable with the typical distance between two close
vortices, is not much smaller than the size of the basin.

\subsection{Random walk on a fractal object: an anomalous diffusive case}
\label{sec:3.3}
In this section we discuss the case of particles performing 
a continuous random walk on a fractal object of fractal dimension $D_F$,
where one has sub-diffusion.
We show that also in situation of anomalous diffusion
(e.g.  sub-diffusion) the FSLE is able to recognize the correct
behavior. In the following section we consider the case of fully
developed turbulence which displays super-diffusion.
 
In a fractal object due to the presence of voids, i.e.
forbidden regions for the particles, one expects a decreasing of the
spreading, and because of the self-similar structure of the
domain (i.e. voids on all scales) a sub-diffusive
behavior is expected.  It is worth to note that the particles do not
diffuse with the same law from any points of the domain (due to the
presence of voids), hence in order to define a diffusive-like behavior 
one has to average over all possible particles' position.
For discrete random walk on a fractal lattice it is known that 
the diffusion follows the law $<R^2(t)> \sim t^{2/D_W}$ with 
$D_W>2$, i.e. sub-diffusion \cite{rammal83}. The quantity $D_W$
is related to the spectral or {\it fracton} dimension $D_S$, by the relation
$D_W=2 D_F/D_S$, and it depends on the detailed structure of the fractal object
\cite{rammal83}.

We study the relative dispersion of 2-D continuous random walk in a
Sierpinsky Carpet with fractal dimension $D_F=\log\,8/\log\,3$.  In our computation 
we use a resolution $3^{-5}$, i.e. the fractal is approximated by five
steps of the recursive building rule, in practice we perform a 
continuous random walk 
in a basin obtained with the above approximation of the Sierpinsky Carpet. 
We initialize the particles inside one of
the smallest resolved structures, then we follow the growth of the relative
dispersion with the FSLE method, and redeploy the
particles in a small cell randomly chosen at the beginning of each
doubling time experiment.  From fig.~4 one can see that
$\lambda(\delta)\sim \delta^{-1/.45}$ which is an indication of
sub-diffusion the exponent is in good agreement with the usual
relative dispersion analysis (see the inset of fig.~4).

\section{Application of the FSLE}
\label{sec:4}

\subsection{Drifter in the Adriatic Sea: data analysis and modelization}
\label{sec:4.1}
Lagrangian data recorded within oceanographic programs in the
Mediterranean Sea \cite{Poulain98} offer the opportunity to apply the
fixed scale analysis to a geophysical problem, for
which the standard characterization of the dispersion properties gives
poor information. 
    
The Adriatic Sea is a semi-enclosed basin, about 800 by 200 $km$ wide, 
connected to the whole Mediterranean Sea through the Otranto Strait  
\cite{Poulain98,ABPPRR97}. We adopt the reference frame in which the $x,y-$axes are 
aligned, respectively, with 
the short side (transverse direction), orthogonal to the coasts, 
and the long side (longitudinal direction), along the coasts.

We have computed relative dispersion along the two axes, 
$\langle R^2_x(t)\rangle$, $\langle R^2_y(t) \rangle$ and 
FSLE $\lambda(\delta)$. 
The number of selected drifters for the analysis is 37, 
distributed in 5 different deployments in the Strait of Otranto, 
happened during the period December 1994 - March 1997, 
containing respectively 4, 9, 7, 7 and 10 drifters. 
To get as high statistics as possible,
even to the cost of losing information on the seasonal variability, 
we shift the time tracks of all of the 37 drifters 
to $t-t_0$, where $t_0$ is the time of the deployment, 
so that the drifters can be treated as a whole cluster. 
Moreover, to restrict the analysis only to the Adriatic basin, we 
discarded a drifter as soon as its latitude 
goes south of $39.5$ N or its longitude goes beyond $19.5$ E.  

Before presenting the results of the data analysis, let us
introduce a simplified model for the Lagrangian tracers motion in the
Adriatic Sea.  We assume as main features of the surface
circulation the following elements \cite{Lacorataetal98}: the drifter
motion is basically two-dimensional; the domain is a quasi-closed
basin; an anti-clockwise coastal current; two large cyclonic gyres;
some natural irregularities in the Lagrangian motion induced by small
scale structures.
On the basis of these considerations, we introduce a 
deterministic chaotic model with mixing properties for the 
Lagrangian drifters. 
The stream function is given by the sum of three terms: 
\beq
\Psi(x,y,t)=\Psi_0(x,y) + \Psi_1(x,y,t) + \Psi_2(x,y,t) \label{4.1} \eeq 
defined
as follows: \beq \Psi_0(x,y)={C_0 \over k_0} \cdot [- \sin(k_0 (y +
\pi)) + \cos(k_0 (x + 2 \pi))] \label{4.2} \eeq 
\beq \Psi_i(x,y,t)={C_i \over k_i }
\cdot \sin(k_i (x + \epsilon_i \sin(\omega_i t))) \sin(k_i (y +
\epsilon_i \sin(\omega_i t + \phi_i))) \,,\hspace{10pt} 
(i=1,2)\,, \label{4.3} 
\eeq 
where $k_i
\,=\, 2 \pi / \lambda_i$, for $i=0,1,2$, $\lambda_i$'s are the
wavelengths of the spatial structures of the flow; analogously
$\omega_j \,=\, 2 \pi / T_j$, for $j=1,2$, and $T_j$'s are the periods
of the perturbations.  In the non-dimensional expression of the
equations, the length and time units have been set to $200 \; km$
and $7.5 \; days$, respectively.

The stationary term 
$\Psi_0$ defines the boundary large scale circulation with positive vorticity. 
The contribution of $\Psi_1$ contains 
the two cyclonic gyres and it is explicitly time-dependent through a periodic 
perturbation. The term $\Psi_2$  
gives the motion over scales smaller than the size of the large gyres 
and it is time-dependent as well.  
The zero-value isoline is defined as the boundary of the basin.  

According to observation, we have chosen the parameters so that  
the velocity range is around  
 $\sim \; 0.3 \; m \, s^{-1}$; 
the length scales of the Eulerian structures  
are $L_B\sim \; 1000 \; km$ (coastal current), 
$L_0\sim \; 200 \; km$ (gyres) and 
$l_u\sim \; 50 \; km$  (vortices); the typical 
recirculation times, for gyres and vortices, 
are $\sim$ 1 month and $\sim$ 1 week, respectively; 
the oscillation periods     
are $ \simeq 10 \, days $ (gyres) and  
$ \simeq 2 \, days$ (vortices). 

Let us discuss now the comparison between data and model results. 
The relative dispersion along the two directions 
of the basin, for data and model trajectories, are shown 
in figures 5a,b. 
The results for the model are obtained from 
the spreading of a cluster of $10^4$ initial conditions. When 
a particle reaches the boundary ($\Psi = 0$) it is eliminated. 

For the diffusion properties, 
one cannot expect a scaling for $\langle R^{2}_{x,y}(t) \rangle$ 
before the saturation regime, since the Eulerian characteristic lengths
are not too small compared with the basin size. Indeed, we do not
observe a power law behavior neither for the experimental data
nor for the numerical model.

Let us stress that  by opportunely fitting the parameters, 
we could obtain the model curves even closer to the experimental ones, 
but this would not be very meaningful since there is no clear theoretical 
expectation in a transient regime. 

Let us now discuss the finite size Lyapunov exponent.  
The analysis of the experimental data has been averaged over the total number 
of couples out of 37 trajectories, under the condition that 
the evolution of the distance between two drifters is no longer followed 
when any of the two exits the Adriatic basin. 

In fig. 6 we show the FSLE for data and model. 
In our case, 
as discussed above, we are far from asymptotical conditions,
therefore we do not observe the scaling $\lambda(\delta) \sim \delta^{-2}$.
 
The $\lambda_{M}(\delta)$ obtained from the minimal chaotic model
(\ref{4.1}-\ref{4.3}) shows the typical step-like shape of a system
with two characteristic time scales, and offers a
scenario about how the FSLE of real trajectories may come out.

The relevant fact is that the large-scale Lagrangian features
are well reproduced, at least at a qualitative level, by a relatively simple model. 
We believe
that this agreement is not due to a particular choice of the model
parameters, but rather to the fact that transport is mainly dominated by
large scales whereas small scale details play a marginal role.

It is  evident the major advantages of FSLE with respect to the
usual fixed time statistics of relative dispersion: 
from the relative dispersion analysis of fig.~5 we are unable to 
recognize the underlying Eulerian structures, while the FSLE
of fig.~6 suggests the presence of structures on different scales and with 
different characteristic times. In conclusion, the fixed scale analysis
gives information for discriminating among different models for the Adriatic Sea.

\subsection{Relative dispersion in fully developed turbulence}
\label{sec:4.2}
We consider now the relative dispersion of particles pairs
advected by an incompressible, homogeneous, isotropic,
fully developed turbulent field.
The Eulerian statistics of velocity differences is characterized by
the Kolmogorov scaling $\delta v(r) \sim r^{1/3}$,
in an interval of scales $\l_u \ll r \ll L_0$, called 
the inertial range, $l_u$ is now the Kolmogorov scale. 
Due to the incompressibility of
the velocity field particles will typically diffuse away from each other
\cite{Cocke69,Orszag70}.
For pair separations less than $l_u$ we have exponential growth of the 
separation of trajectories, typical of smooth flows,
whereas at separations larger than $L_0$ normal diffusion takes place.
In the inertial range the average pair separation 
is not affected neither by large scale components of the flow,
which simply sweep the pair, nor by small scale ones, whose intensity is
low and which act incoherently.
Accordingly, the separation $R(t)$ feels mainly the 
action of velocity differences $\delta v(R(t))$ at scale $R$.
As a consequence of the Kolmogorov scaling the separation grows with
the {\em Richardson law} \cite{Richardson26,MY75}
\beq
\langle R^2(t) \rangle  \sim t^{3} \, .
\label{3.2.1}
\eeq    

Non-asymptotic behavior takes place in such systems
whenever $l_u$ is not much smaller than $L_0$, that is
when the Reynolds number is not high enough.
As a matter of facts
even at very high Reynolds numbers, the inertial range is
still insufficient to observe the scaling (\ref{3.2.1}) without any ambiguity. 
On the other hand, we shall
show that FSLE statistics is effective already a relatively small 
Reynolds numbers.

In order to investigate the problem of relative dispersion at various
scale separations a practical tool is the use of synthetic turbulent
fields. In fact, by means of stochastic processes it is possible to
build a velocity field which reproduces the statistical properties of
velocity differences observed in fully developed turbulence
\cite{BBCCV98}. In order to avoid the difficulties related to the
presence of sweeping in the velocity field, we limit ourselves to a
correct representation of two-point velocity differences. In this
case, if one adopts the reference frame in which one of the two
tracers is at rest at the origin (the so called a Quasi-Lagrangian
frame of reference), the motion of the second particle is ruled
by the velocity difference in this frame of reference, which has the
the same single time statistics of the Eulerian
velocity differences \cite{LPP97,BCCV99}.  The detailed construction
of the synthetic Quasi-Lagrangian velocity field is presented in
Appendix \ref{app:3}.

In figure 7
we show the results of simulations of pair dispersion 
by the synthetic turbulent field with Kolmogorov scaling 
of velocity differences at Reynolds number $Re \simeq 10^{6}$ \cite{BCCV99}.
The expected  super-diffusive regime (\ref{3.2.1}) can be 
well observed only for huge Reynolds numbers (see also Ref. \cite{Maida}.
To explain the depletion of scaling range for the relative dispersion
let us consider a series of pair dispersion
experiments, in which a couple of particles is released at a
separation $R_0$ at time $t=0$. At a fixed time $t$,
as customarily is done, we perform an average over all different
experiments to compute $\langle R^2(t) \rangle$.
But, unless $t$ is large enough that all particle pairs have
``forgotten'' their initial conditions, the average will be biased.
This is at the origin of the flattening of $\langle R^2(t) \rangle$
for small times, which we can call a crossover from initial
condition to self similarity. In an analogous fashion
there is a crossover for large times,
of the order of the integral time-scale, since some couples might have
reached a separation larger than the integral scale, and thus
diffuse normally, meanwhile other pairs still lie within the inertial
range, biasing the average and, again, flattening the curve
$\langle R^2(t) \rangle$.
This correction to a pure power law is far from being negligible for
instance in experimental data where the inertial range is generally
limited due to the Reynolds number and the experimental apparatus.
For example, references \cite{FV98,Fung92} show quite clearly the
difficulties that may arise in numerical simulations with the standard
approach.

To overcome these difficulties we exploit the  approach 
based on the fixed scale statistics.
The outstanding advantage of averaging at a fixed separation scale
is that it removes all crossover effects, since all sampled pairs
belong to the inertial range.
The expected scaling properties of the doubling times is obtained by a
simple dimensional argument. The time it takes for particle separation
to grow from $R$ to $2 R$ can be estimate as $T(R) \sim R/\delta v(R)$;
we thus expect for the inverse doubling times the scaling
\begin{equation}
\left\langle {1 \over T(R)} \right\rangle \sim
       {\langle \delta v(R) \rangle \over R}
\sim R^{-2/3}
\label{eq:5.1}
\end{equation}
In figure 8 the great enhancement of the scaling range
achieved by using the doubling times is evident.  In addition,
by using the FSLE it is possible to study in details the effect of Eulerian
intermittency on the Lagrangian statistics of relative dispersion.
See Ref.\cite{BCCV99} for a detailed discussion and a comparison with a
multifractal scenario.  The conclusion that can be drawn 
is that in this case  doubling time statistics makes it possible  a much
better estimate of the scaling exponent with respect to the standard --
fixed time -- statistics.

\section{Conclusions}
\label{sec:5}
In the study of relative dispersion of Lagrangian 
tracers one has to tackle situations in which the asymptotic 
behavior is never attained. This may happen 
in presence of many characteristic Eulerian scales or, what is
typical of real systems, in presence of boundaries.
It is worth to stress that such kind of systems are very common
in geophysical flows \cite{D}, and also in plasma physics \cite{sabot}.
Therefore a close understanding of non-asymptotic transport properties 
can give much relevant information about these natural phenomena.

To face these problems, in recent years, there have been proposed
different approaches whose common ingredient is basically an ``exit
time'' analysis. We remind the symbolic dynamics \cite{A,B} and 
the chaotic scattering \cite{C} approaches, the exit time description 
for transport in semi-enclosed basins \cite{D}, symplectic maps \cite{meiss},
open flows \cite{E} and in plasma physics \cite{sabot}. 

In this paper we have discussed the applications of the Finite Size 
Lyapunov Exponent, $\lambda(\delta)$, in the analysis of several situations.
This method is based on the identification of the typical
time $\tau(\delta)$ characterizing the diffusive process at scale
$\delta$ through the exit time.
This approach is complementary to the traditional one, in which one looks
at the average size of the clouds of tracers as function of time.
For values of $\delta$  much smaller than
the smallest characteristic length of the Eulerian velocity field, one
has that $\lambda(\delta)$ coincides with the maximum Lagrangian
Lyapunov Exponent.  For larger $\delta$ the shape of $\lambda(\delta)$
depends on the detailed mechanisms of spreading, i.e. the structure of
the advecting velocity field and/or the presence of boundaries.
The diffusive regime corresponds to the behavior
$\lambda(\delta)\simeq D/\delta^2$. If $\delta$ gets close to its
saturation value, i.e. the characteristic size of the basin, the
universal shape of $\lambda(\delta)$ can be obtained on the basis of
dynamical system theory.  In addition, we have shown that 
the fixed scale method is able to recognize the presence of a genuine 
anomalous diffusion.

A remarkable advantage of working at fixed scale (instead of at fixed
time as in the traditional approach) is its ability to avoid
misleading results, for instance apparent anomalous scaling over a
certain time interval. Moreover, with the FSLE one obtains the proper
scaling laws also for a relatively small inertial range for which the
standard technique gives rather controversial answers.

The proposed method can be also applied in the analysis of drifter
experimental data or in numerical model for Lagrangian transport.

\section{Acknowledgments}
We thank V. Artale, E. Aurell, L. Biferale, P. Castiglione, A. Crisanti, 
M. Falcioni, R. Pasmanter, P.M. Poulain, M. Vergassola and E. Zambianchi
for collaborations and discussions in last years.  
A particular acknowledgment to B. Marani for the
continuous and warm encouragement.  We are grateful to the ESF-TAO
({\it Transport Processes in the Atmosphere and the Oceans})
Scientific Program for providing meeting opportunities.  This paper
has been partly supported by INFM (Progetto di Ricerca Avanzato
PRA-TURBO), MURST (no. 9702265437), and the European Network {\it
Intermittency in Turbulent Systems} (contract number FMRX-CT98-0175).

\appendix

\section{Computation of the Finite size Lyapunov exponent}
\label{app:1}

In this appendix we discuss in detail the method for computing the Finite
Size Lyapunov Exponent for both continuous dynamics (differential
equations) and discrete dynamics (maps).

The practical method for computing the FSLE goes as follows.
Defined a given norm for the distance $\delta(t)$ between the
reference and perturbed trajectories, one has to define a series of thresholds
$\delta_{n}=r^{n}\delta_{0}$ ($n=1,\dots, P$),
and to measure the ``doubling times'' $T_{r}(\delta_{n})$ that a
perturbation of
size $\delta_{n}$ takes to grow up to $\delta_{n+1}$.
The threshold rate $r$ should not be taken too large, because
otherwise the error has to grow through different scales before
reaching the next threshold. On the other hand, $r$ cannot be
too close to one, because otherwise the doubling time would be
of the order of the time step in the integration. In our examples
we typically use $r=2$ or $r=\sqrt 2$. For simplicity $T_{r}$
is called ``doubling time'' even if $r \neq 2$.

The doubling times $T_{r}(\delta_{n})$ are obtained by following
the evolution of the
separation from its initial size $\delta_{min} \ll \delta_0$ up to
the largest threshold $\delta_{P}$.
This is done by
integrating the two trajectories of the system starting at an initial
distance $\delta_{min}$. In general, one must choose
$\delta_{min} \ll \delta_{0}$, in order to allow the direction of the
initial perturbation to align with the most unstable direction in the
phase-space. Moreover, one must pay attention to keep
$\delta_{P} < \delta_{max}$, so that all the thresholds
can be attained ($\delta_{max}$ is the typical distance
of two uncorrelated trajectory).

The evolution of the error from the initial value $\delta_{min}$ to
the largest threshold $\delta_{P}$ carries out a single error-doubling
experiment. At this point one rescales the model trajectory at the
initial distance $\delta_{min}$ with respect to the true trajectory
and starts another experiment.
After ${N}$ error-doubling experiments, we can estimate the
expectation value of some quantity $A$ as:
\begin{equation}
\langle A \rangle_{e} = {1 \over {N}} \sum_{i=1}^{N} \, A_i \, .
\label{eq:ap1}
\end{equation}
This is not the same as taking the time average 
because different error doubling experiments may takes different times.
Indeed we have
\begin{equation}
\langle A \rangle_{t} = {1 \over T} \int_0^T \, A(t) dt =
{\sum_i A_i \tau_i \over \sum_i \tau_i} =
{\langle A \tau \rangle_{e} \over \langle \tau \rangle_{e}} \, .
\label{eq:ap2}
\end{equation}
In the particular case in which $A$ is the doubling time itself we have
from (\ref{eq:ap2})
\begin{equation}
\lambda(\delta_n) = {1 \over \langle T_{r}(\delta_n) \rangle_{e}} \ln r \, .
\label{eq:ap3}
\end{equation}

The method above described assumes that the distance between the two
trajectories is continuous in time. This is not true for maps or for
discrete sampling in time, thus the method has to be slightly modified.
In this case $T_{r}(\delta_n)$ is defined as the minimum time at which
$\delta(T_r) \ge r \delta_n$. Because now $\delta(T_r)$ is a fluctuating
quantity, from (\ref{eq:ap2}) we have
\begin{equation}
\lambda(\delta_n) = {1 \over \langle T_{r}(\delta_n) \rangle_{e}}
\left\langle \ln \left( {\delta(T_r) \over \delta_n} \right) 
\right\rangle_{e} \,.
\label{eq:ap4}
\end{equation}

We conclude by observing that the computation of the FSLE is not more
expensive than the computation of the Lyapunov exponent by standard
algorithm. One has simply to integrate two copies of the system and
this can be done also for very complex simulations.

\section{Universal saturation behavior of $\lambda(\delta)$}
\label{app:2}
In this appendix we present the derivation of the 
asymptotic behavior (\ref{eq:nearbound}) of $\lambda(\delta)$
for $\delta$ close to the saturation. The computation is
explicitly done for the simple case of a one dimensional Brownian
motion in the domain  $[-L_B,L_B]$, with reflecting boundary conditions:
the numerical simulations indicate that the result is of general 
applicability.

The evolution of the probability density $p$ is ruled by the 
Fokker-Planck equation
\begin{equation}
{\partial p \over \partial t} = {1 \over 2} D
 {\partial^2 p \over \partial x^2}
\label{FP}
\end{equation}
with the Neumann boundary conditions
${\partial p \over \partial x}(\pm L_B)=0$ .

The general solution of (\ref{FP}) is 
\begin{equation}
p(x,t)= {\displaystyle \sum_{k=-\infty}^{\infty} \hat{p}(k,0) e^{i k x}
e^{-t/ \tau_k} + c.c }\\  
\label{sol}
\end{equation}
where 
\begin{equation}
{\displaystyle\tau_k=\left( {D \over 2} {\pi^2 \over L_B^2} k^2  \right)^{-1}} ,
\, \, \, k=0, \pm 1, \pm 2, ...
\end{equation}
At large times $p$ approaches the uniform solution 
$p_0=1/2L_B$.
Writing $p$ as  $p(x,t)= p_0 + \delta p(x,t)$
we have, for $t \gg \tau_1$ ,
\begin{equation}
\delta p \sim \exp(-t/\tau_1).
\label{asi}
\end{equation} 
The asymptotic behavior for the relative dispersion
$\langle R^2(t) \rangle$ is
\begin{equation}
\langle R^2(t) \rangle={1 \over 2} \int (x-x')^2 p(x,t) p(x',t) \;dx\; dx' 
\end{equation}
For $t \gg \tau_1$ using (\ref{asi}) we obtain
$\langle R^2(t)\rangle  \sim \left( {L_B^2 \over 3}-A e^{-t/ \tau_1} \right)$.
Therefore for  $\delta^2(t)=\langle R^2(t) \rangle$ one has
$\delta(t) \sim 
\left( {L_B \over \sqrt{3}} -{\sqrt{3} A \over 2 L_B}  e^{-t/ \tau1} \right)$
The saturation value of $\delta$ is 
$\delta_{max}=L_B/\sqrt{3}$, so 
for $t \gg \tau_1$, or equivalently for 
$(\delta_{max}-\delta)/\delta \ll 1$, we expect
\begin{equation}
{d \over dt} \ln \delta =\lambda({\delta})=
{1 \over \tau_1}{\delta_{max}-\delta \over \delta}
\label{satu}
\end{equation}
which is (\ref{eq:nearbound}).

Let us remark that in the previous argument for 
$\lambda(\delta)$ for
$\delta \simeq \delta_{max}$ the crucial point is the 
exponential relaxation to the asymptotic  uniform distribution.
In a generic deterministic chaotic system it is not
possible to prove this property in a rigorous way.
Nevertheless one can expect that this request is fulfilled 
at least in  non-pathological cases. In  
chaotic systems the exponential relaxation to asymptotic 
distribution corresponds to have the second eigenvalue $\alpha$ of the 
Perron-Frobenius operator inside the unitary circle;
the relaxation time is $\tau_1=-\ln|\alpha|$ \cite{Beck}.

\section{Synthetic turbulent velocity fields}
\label{app:3}
The generation of a synthetic turbulent field which reproduces
the relevant statistical features of fully developed
turbulence is not an easy task. Indeed to obtain a physically
sensible evolution for the velocity field one has to take into account
the fact that each eddy is subject to the action of all
other eddies. Actually the overall effect amounts only
to two main contributions, namely the
sweeping exerted by larger eddies and
the shearing due to eddies of comparable size.
This is indeed a substantial simplification, but nevertheless
the problem of properly mimicking the effect of sweeping
is still unsolved.
                        
To get rid of these difficulties
we shall limit ourselves to the generation of a synthetic
velocity field in {\em Quasi-Lagrangian} (QL) coordinates \cite{LPP97},
thus moving to a frame of reference
attached to a particle of fluid $\bbox{r}_{1}(t)$.
This choice bypasses the problem of sweeping, since
it allows to work only with relative velocities,
unaffected by advection.
As a matter of fact there is a price to pay for the considerable
advantage gained by discarding advection, and it is that
only the problem of two-particle
dispersion can be well managed within this framework.
The properties of single-particle Lagrangian statistics cannot,
on the contrary, be consistently treated.

The QL velocity differences
are defined as
\begin{equation}
\bbox{v}(\bbox{r},t) = \bbox{u}\left(\bbox{r}_{1}(t)+\bbox{r},t\right)
- \bbox{u}\left(\bbox{r}_{1}(t),t\right) \; ,
\label{eq:2.1}
\end{equation}

where the reference particle moves according to
\begin{equation}
{{\rm d}\bbox{r}_{1}(t) \over {\rm d}t} = \bbox{u}(\bbox{r}_{1}(t),t) \; .
\label{eq:2.2}
\end{equation}
These  velocity differences have the useful property
that their single-time statistics are the same as the Eulerian ones
whenever considering statistically stationary flows \cite{LPP97}.
For fully developed turbulent flows,
in the inertial interval of length scales where
both viscosity and  forcing are negligible, the QL longitudinal
velocity differences show the scaling behavior
\begin{equation}
\langle \left| \bbox{v}(\bbox{r}) \cdot
\frac{\bbox{r}}{r} \right|^p \rangle \sim r^{\zeta_p}
\label{eq:2.3bis}
\end{equation}
where the exponent $\zeta_p$ is a convex function of $p$,
and $\zeta_3=1$.
This scaling behavior is a distinctive
statistical property of fully developed turbulent flows
that we shall reproduce by means of a synthetic velocity field.
In the QL reference frame the first particle is at rest in the origin
and the second particle is at
$\bbox{r}_{2}=\bbox{r}_{1}+\bbox{R}$, advected with respect to the
reference particle by the relative velocity
\begin{equation}
\bbox{v}(\bbox{R},t) =
\bbox{u}\left(\bbox{r}_{1}(t)+\bbox{R},t\right) -
\bbox{u}\left(\bbox{r}_{1}(t),t\right)
\label{eq:2.3}
\end{equation}
By this change of coordinates
the problem of pair dispersion in an Eulerian velocity field has
been reduced to the problem of single particle dispersion in
the velocity difference field $\bbox{v}(\bbox{r},t)$.
This yields a substantial simplification:
it is indeed sufficient to build a velocity difference field
with proper scaling features in the radial direction only,
that is along the line that joins the reference particle
$\bbox{r}_{1}(t)$ -- at rest in the origin of the QL coordinates --
to the second particle $\bbox{r}_{2}(t)=\bbox{r}_{1}(t)+\bbox{R}(t)$.
To appreciate this simplification, it must be noted that actually
all moments of velocity differences
$\bbox{u}\left(\bbox{r}_{1}(t)+\bbox{r}',t\right) -
\bbox{u}\left(\bbox{r}_{1}(t)+\bbox{r},t\right)=
\bbox{v}(\bbox{r}',t)-\bbox{v}(\bbox{r},t)$ should display
power law scaling in
$|\bbox{r}'-\bbox{r}|$. Actually these latter differences
never appear in the dynamics
of pair separation, and so we can limit ourselves to fulfill
the weaker request (\ref{eq:2.3bis}).
Needless to say, already for three particle dispersion one needs
a field with proper scaling in all directions.

We limit ourselves to the two-dimensional case,
where we can introduce a stream function for the QL
velocity differences
\begin{equation}
\bbox{v}(\bbox{r},t) = \nabla \times \psi(\bbox{r},t) \; .
\label{eq:2.4}
\end{equation}
The extension to a three
dimensional velocity field is not difficult but more expensive
in terms of numerical resources.

Under isotropic conditions, the stream function
can be decomposed in radial octaves as
\begin{equation}
\psi(\bbox{r},\theta,t) = \sum_{i=1}^{N}\sum_{j=1}^{n}
            \frac{\phi_{i,j}(t)}{k_i} F(k_i r) G_{i,j}(\theta) \; .
\label{eq:2.5}
\end{equation}
where $k_i=2^{i}$.
Following a heuristic argument, one expects that at a given $r$
the stream function is essentially dominated by the contribution
from the $i$ term such that $r \sim 2^{-i}$.
This locality of contributions suggests a simple choice for
the functional dependencies of the ``basis functions'':
\begin{equation}
F(x) = x^2(1-x)\ \,\,\, \mbox{\rm for}\ \,\, 0\le x \le 1
\end{equation}
and zero otherwise,
\begin{equation}
G_{i,1}(\theta) = 1, \qquad G_{i,2}(\theta) = \cos(2 \theta + \varphi_i)
\end{equation}
and $G_{i,j}=0$ for $j>2$ ($\varphi_i$ is a quenched random phase).
It is worth remarking that this choice is rather general
because it can be derived from the lowest order expansion for small $r$
of a generic streamfunction in Quasi-Lagrangian coordinates.

It is easy to show that, under the usual locality conditions for infra
red convergence, $\zeta_p <p$ \cite{RS78}, the leading contribution to
the $p$-th order longitudinal structure function $\langle |v_r(r)|^p
\rangle$ stems from $M$-th term in the sum (\ref{eq:2.5}), $\langle
|v_r(r)|^p \rangle \sim \langle |\phi_{M,2}|^p \rangle$ with $r \simeq
2^{-M}$.  If the $\phi_{i,j}(t)$ are stochastic processes with
characteristic times $\tau_i=2^{-2i/3}\,\tau_0$, zero mean and
$\langle |\phi_{i,j}|^p\rangle \sim k_i^{-\zeta_p}$, the scaling
(\ref{eq:2.3bis}) will be accomplished.  An efficient way of to
generate $\phi_{i,j}$ is \cite{BBCCV98}:
\begin{equation}
\phi_{i,j}(t) = g_{i,j}(t)\, z_{1,j}(t)\,z_{2,j}(t)\cdots z_{i,j}(t)
\end{equation}
where the $z_{k,j}$ are independent, positive definite, identically
distributed random processes with characteristic time $\tau_k$, while
the $g_{i,j}$ are independent stochastic processes with zero mean,
$\langle g_{i,j}^2\rangle \sim k_i^{-2/3}$ and characteristic time
$\tau_i$.
The scaling exponents $\zeta_p$ are determined by the probability
distribution of $z_{i,j}$ via
\begin{equation}
\label{eq:zetap}
 \zeta_p = \frac{p}{3} - \log_2\langle z^p\rangle \; .
\end{equation}
As a last remark we note that by simply fixing the $z_{i,j}=1$
we recover the Kolmogorov scaling, which has been used in the simulations
presented in section \ref{sec:4.2}
                                        


\newpage
\centerline{\bf FIGURE CAPTIONS}
\begin{description}
\item[Figure 1:]
a) Three realizations of $R^2(t)$ as a function of $t$ built 
as follows:
$R^2(t)=\delta_0^2 \exp(2 \gamma t)$ if $R^2(t)<1$ and 
$R^2(t)=2 D(t-t_{\ast})$ with $\gamma=0.08, 0.05, 0.3$ and 
$\delta_0=10^{-7}, \,\; D=1.5$.
b) $\langle R^2(t) \rangle$ as function of $t$ averaged on 
the three realizations shown in figure 1a. The apparent anomalous
regime and the diffusive one are shown. 
c) $\lambda(\delta)$ vs $\delta$, with Lyapunov and 
diffusive regimes.

\item[Figure 2:]
Lagrangian motion given by the Rayleigh-B\'enard convection model
with: $A=0.2, \,\; B=0.4, \,\; \omega=0.4, \,\; k=1.0, \,\; a=\pi$, 
the number of realizations is ${\cal N}=2000$ and the series of thresholds 
is $\delta_n=\delta_0 r^n$ with $\delta_0=10^{-4}$ and $r=1.05$. 
$\lambda(\delta)$ vs $\delta$, 
in a closed domain with $6$ (crosses) and $12$ (diamonds) 
convective cells. 
The lines are respectively:
(a) Lyapunov regime with $\lambda=0.017$; (b) diffusive regime 
with $D=0.021$; (c) saturation regime with $\delta_{max}=19.7$; 
(d) saturation regime with $\delta_{max}=5.7$.

\item[Figure 3:]
(a)$\langle R^{2}(t) \rangle$ for the four vortex system
with $\Gamma_{1}=\Gamma_{2}=-\Gamma_{3}=-\Gamma_{4}=1$. The 
threshold parameter is $r=1.03$ and $\delta_{0}=10^{-4}$, 
the dashed line is the power law
$\langle R^{2}(t) \rangle \sim t^{1.8}$. The number of 
realizations is ${\cal N} = 2000$.
(b)$\lambda(\delta)$ vs $\delta$ for the same model 
and parameters. The horizontal line indicates 
the Lyapunov exponent ($\lambda=0.14$), 
the dashed curve is the saturation regime with $\delta_{max}=0.76$

\item[Figure 4:]FSLE computed for particle diffusion in a Sierpinsky
Carpet of fractal dimension $D_f=\log(8)/\log(3)$ obtained by iteration
of the unit structure up to a resolution $3^{-5}$, one has:
$\lambda(\delta)\sim \delta^{-1/.45}$, which is
in agreement with the value obtained for $<R(t)>$ versus $t$ shown in the 
inset (i.e. $<R(t)>\sim t^{.45}$).

\item[Figure 5:]
Relative dispersion of Lagrangian trajectories, $<R_{x,y}^2(t)>$ {\it versus} $t$, in 
the Adriatic Sea, for data (continuous line) and model (dashed line), 
along the natural axes of the basin:  
(a) transverse direction ($x-$axis) and  
(b) longitudinal direction ($y-$axis). The time is measured in $days$ and the 
mean square radius of the cluster is in $km^2$.

\item[Figure 6:]
FSLE of Lagrangian trajectories in the Adriatic Sea, for  
data (continuous line) and model (dashed line). The scale 
$\delta$ is in $km$, $\lambda(\delta)$ is in $days^{-1}$. 

\item[Figure 7:]
Relative dispersion $\langle R(t) \rangle$ for $N=20$ octaves
synthetic turbulent simulation averaged over $10^4$ realizations. 
The line is the theoretical Richardson scaling $t^{3/2}$.

\item[Figure 8:]
Average inverse doubling time $\langle 1/T(R) \rangle$ for
the same simulation of the previous figure.
Observe the enhanced scaling region. 
The line is the theoretical Richardson scaling
$R^{-2/3}$.

\end{description}

\newpage
\begin{figure}[p]
\centerline{\epsfig{figure=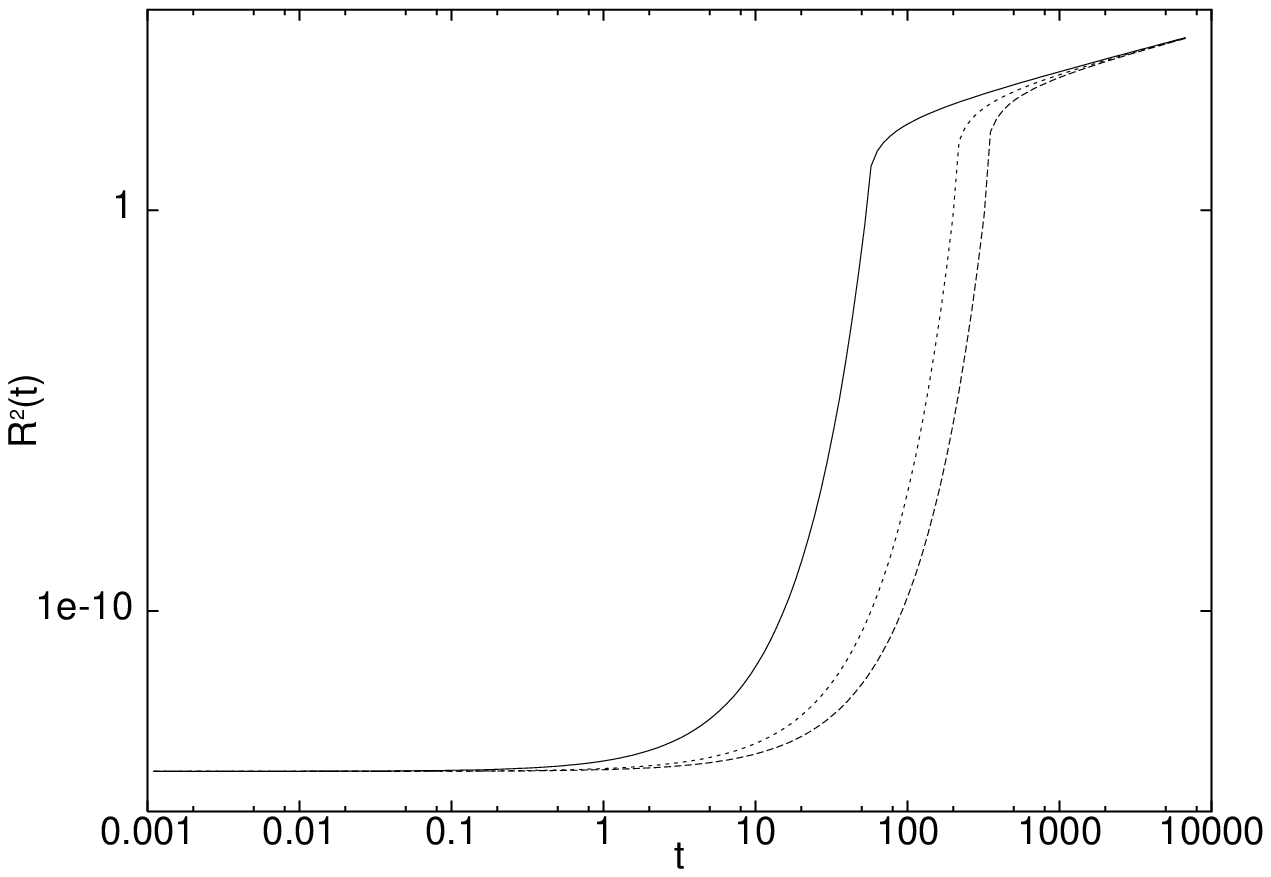,width=7cm,angle=0}}
\vspace{.2cm}
\centerline{\hspace{.5cm} (a)}
\vspace{.8cm}
\centerline{\epsfig{figure=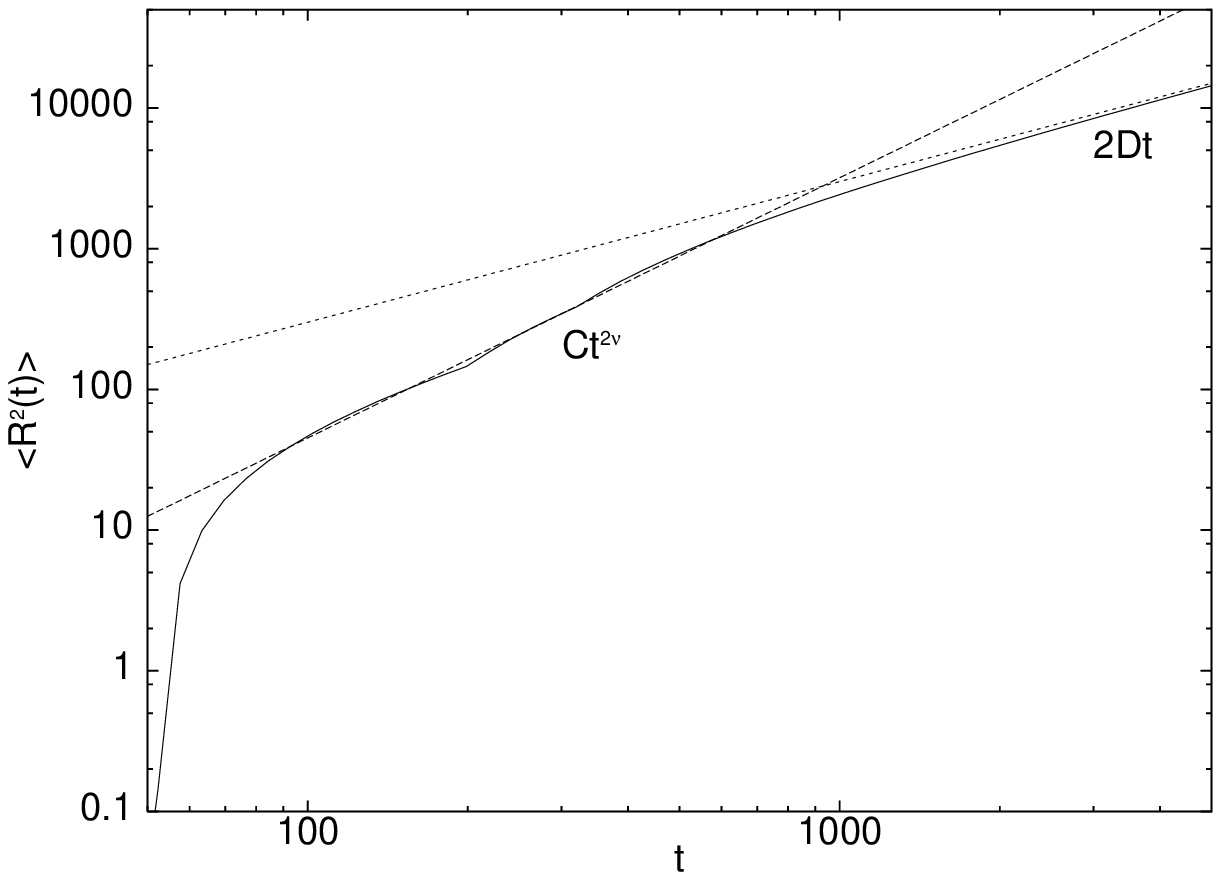,width=7cm,angle=0}}
\vspace{.2cm}
\centerline{\hspace{.5cm} (b)}
\vspace{.8cm}
\centerline{\epsfig{figure=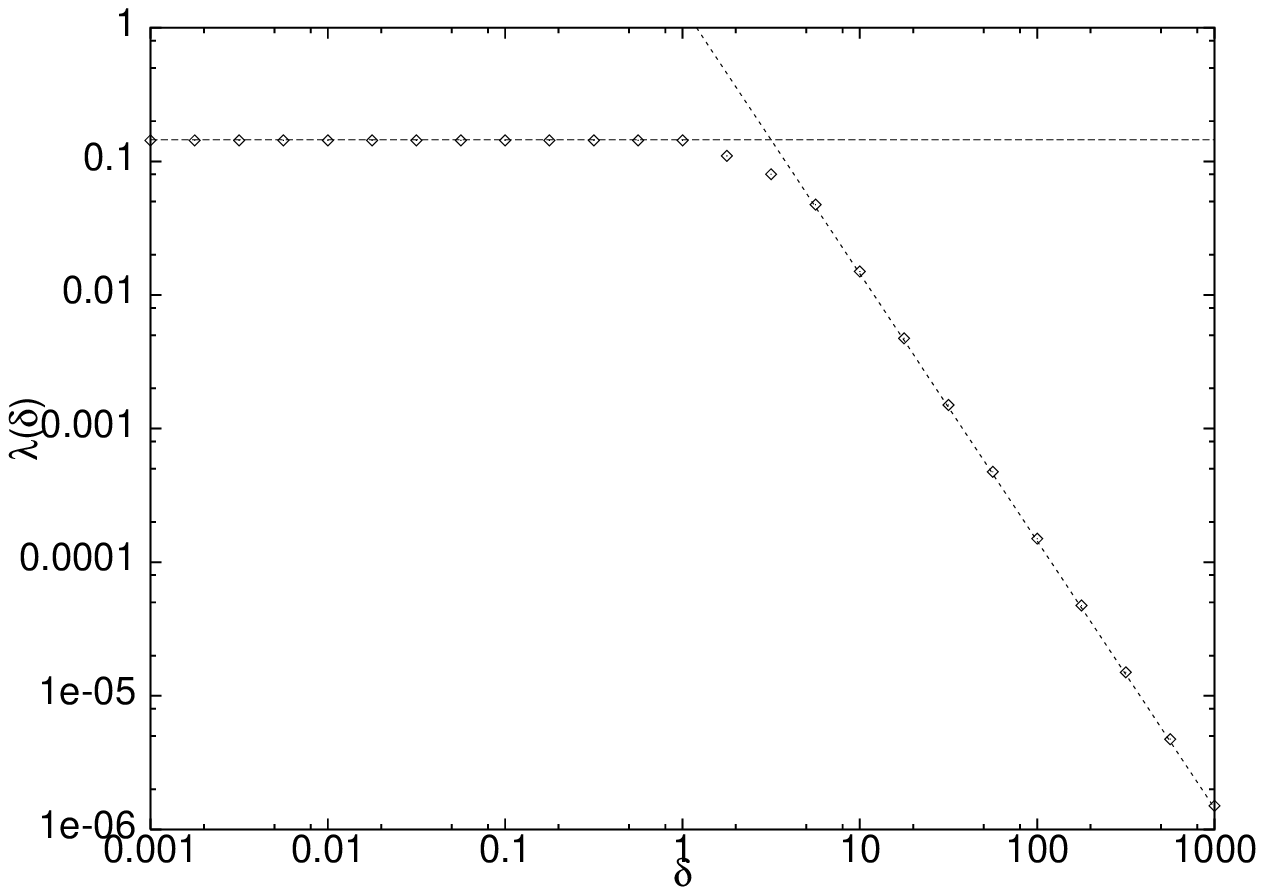,width=7cm,angle=0}}
\vspace{.2cm}
\centerline{\hspace{.5cm} (c)}
\caption{
}  
\label{fig:1}
\end{figure}

\newpage

\begin{figure}[p]
\centerline{\epsfig{figure=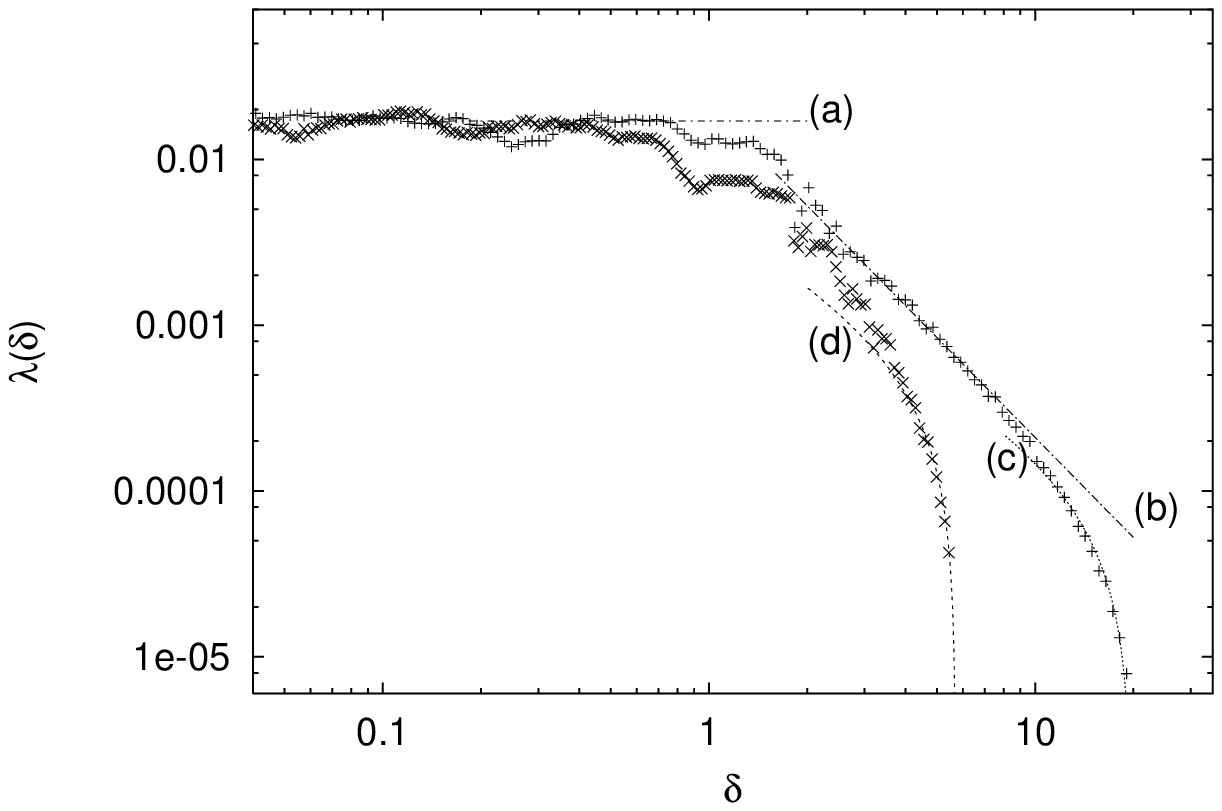,width=9cm,angle=0}}
\caption{}
\label{fig:2}
\end{figure}

\newpage
\begin{figure}[p]
\centerline{\epsfig{figure=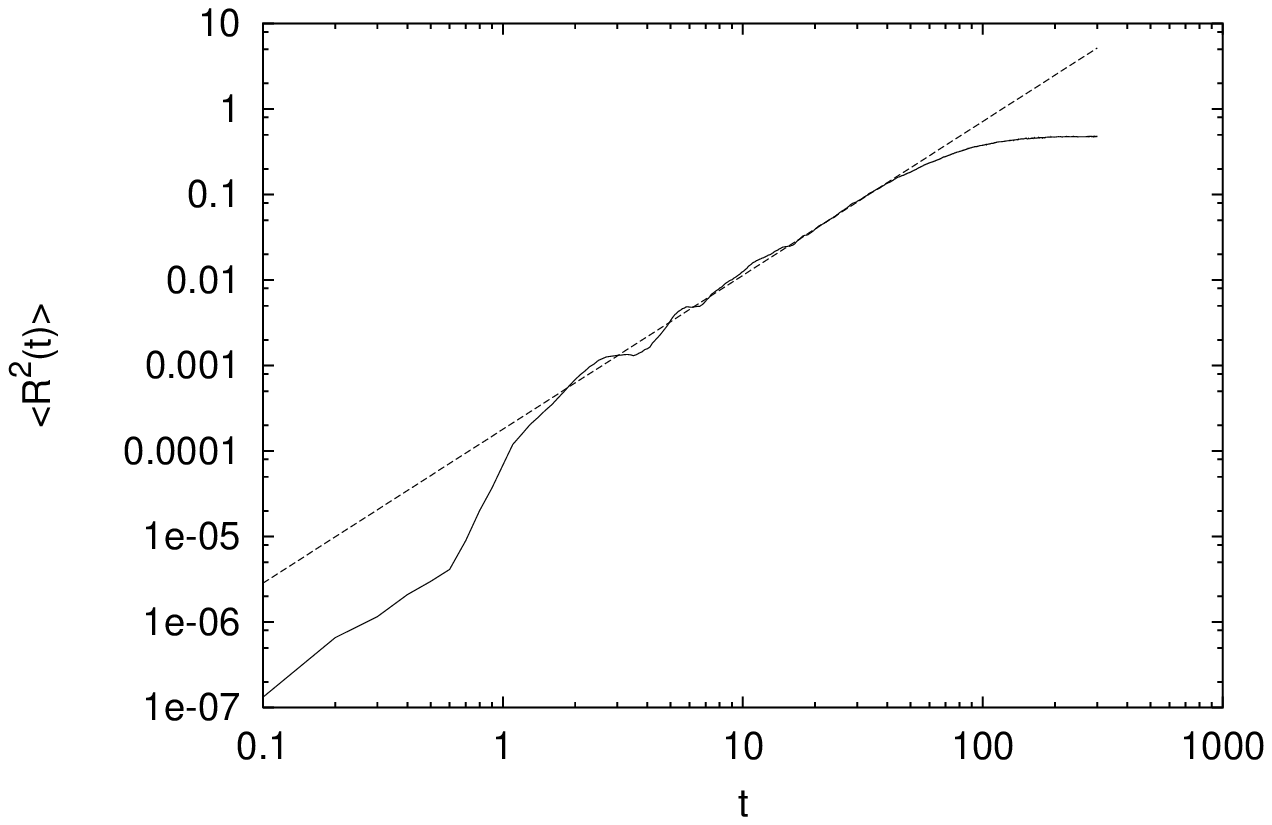,width=9cm,angle=0}}
\centerline{\hspace{1cm} (a)}
\vspace{2.0cm}
\centerline{\epsfig{figure=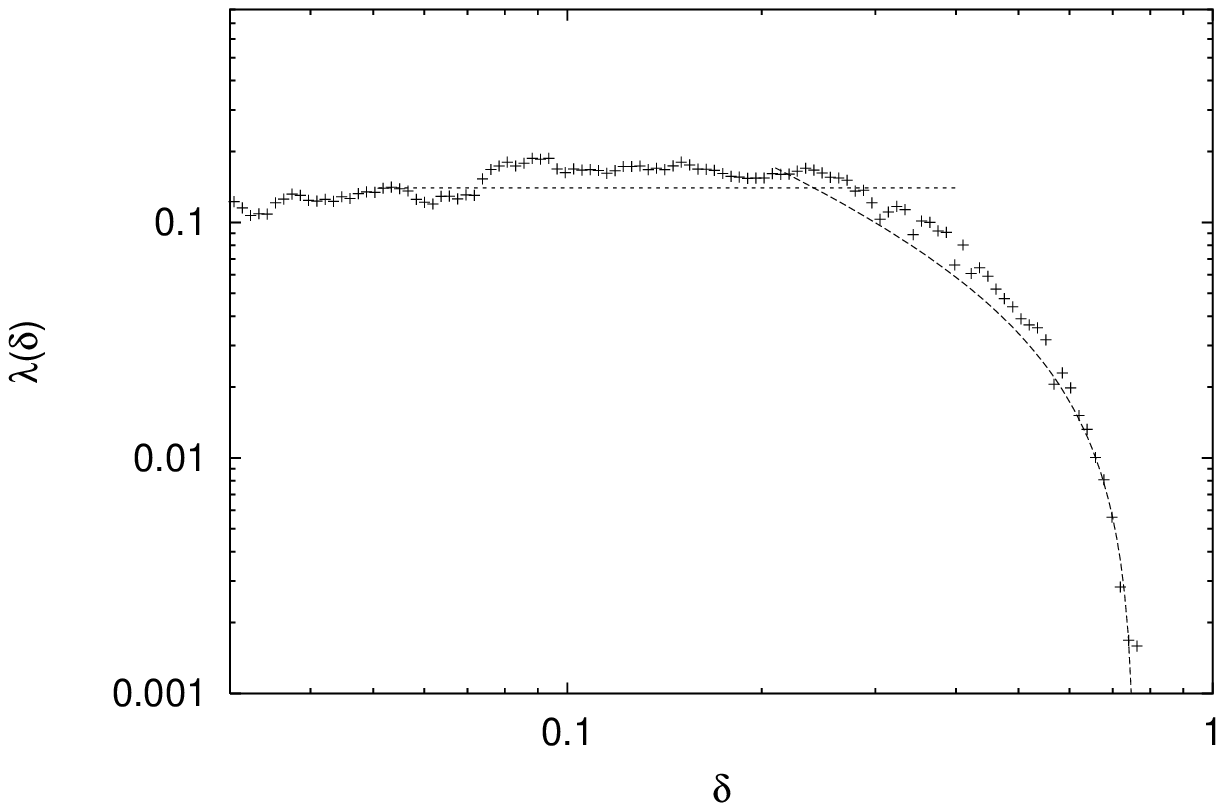,width=9cm,angle=0}} 
\centerline{\hspace{1cm} (b)}
\caption{ }
\label{fig:3}
\end{figure}

\newpage

\begin{figure}[p]
\centerline{\epsfig{figure=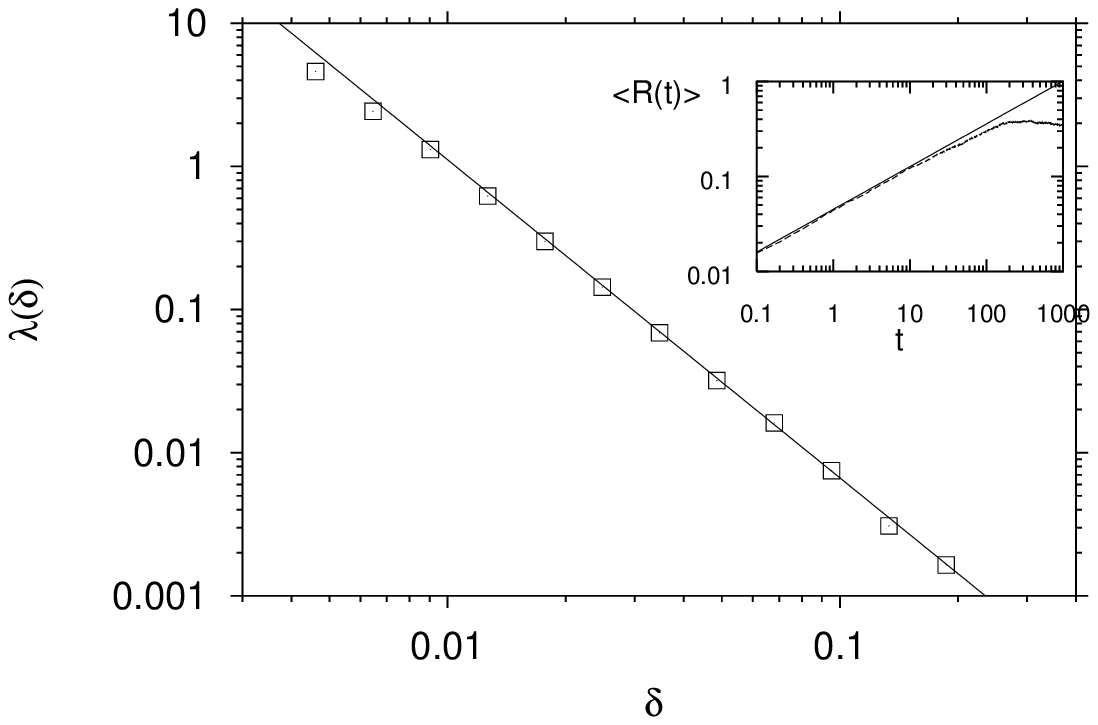,width=9cm,angle=0}}
\caption{}
\label{fig:4}
\end{figure}

\newpage

\begin{figure}[p]
\centerline{\epsfig{figure=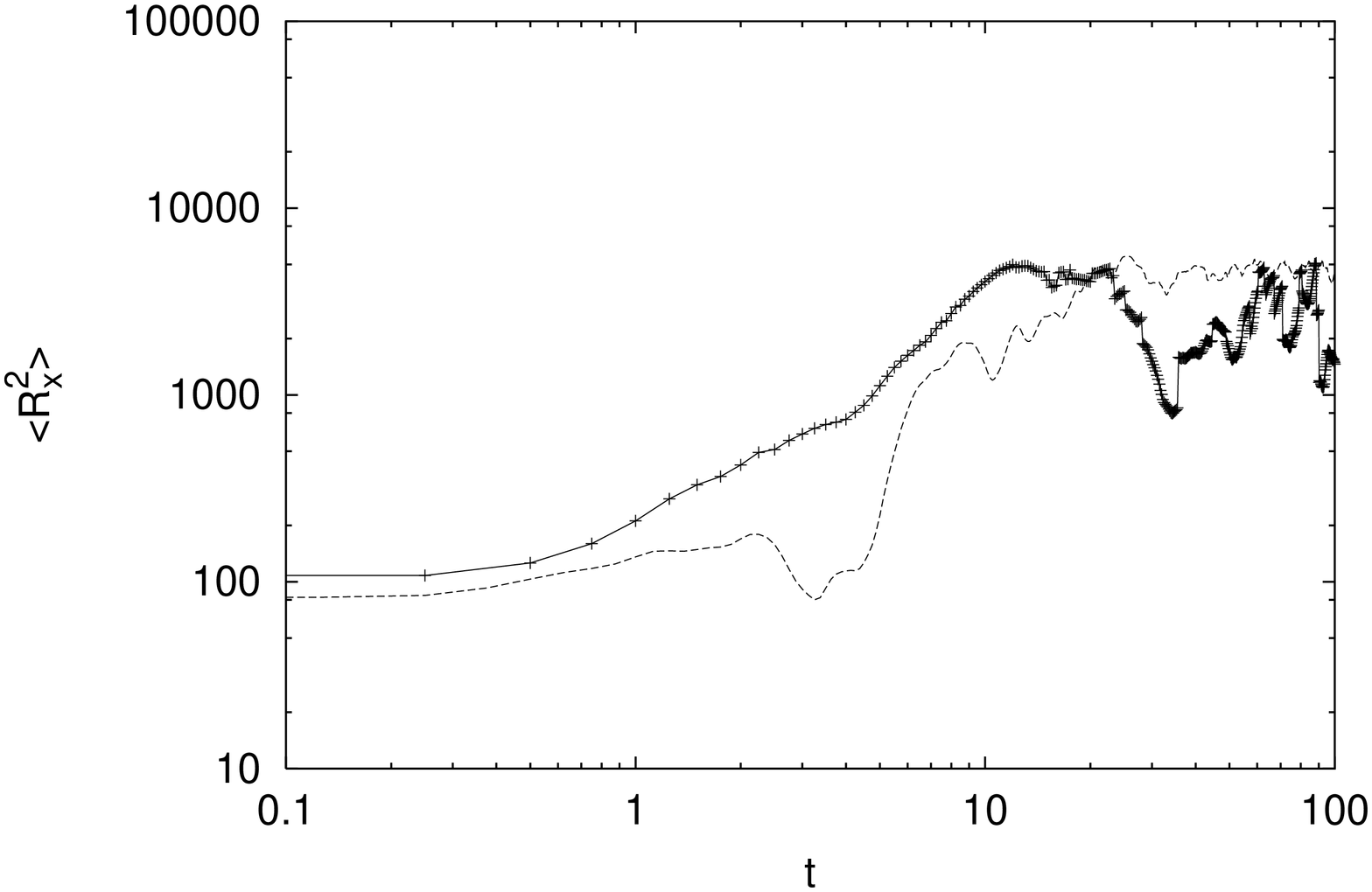,width=6cm,angle=270}}
\centerline{\hspace{1cm} (a)}
\vspace{2.0cm}
\centerline{\epsfig{figure=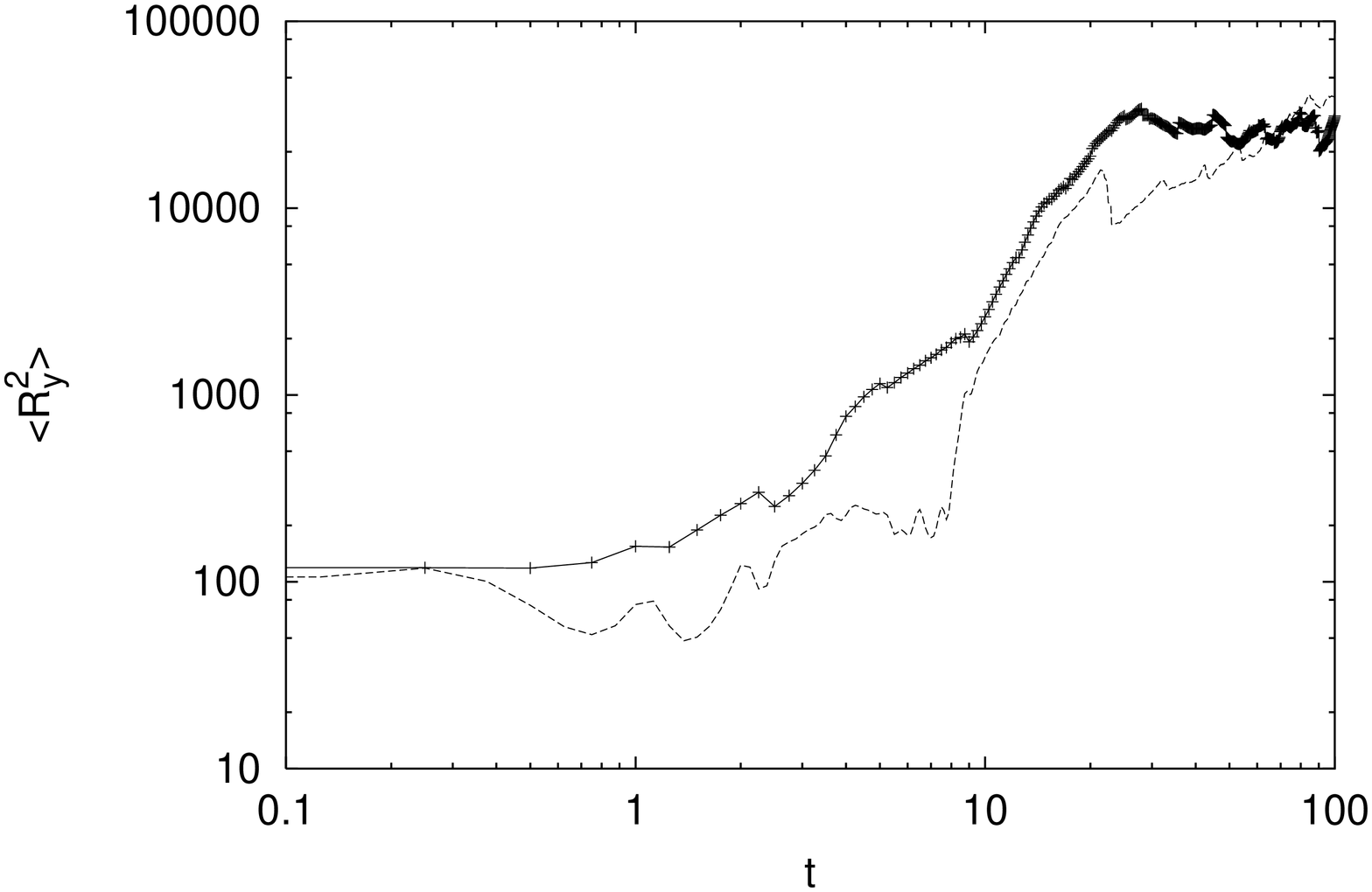,width=6cm,angle=270}} 
\centerline{\hspace{1cm} (b)}
\caption{ }
\label{fig:5}
\end{figure}

\newpage

\begin{figure}[p]
\centerline{\epsfig{figure=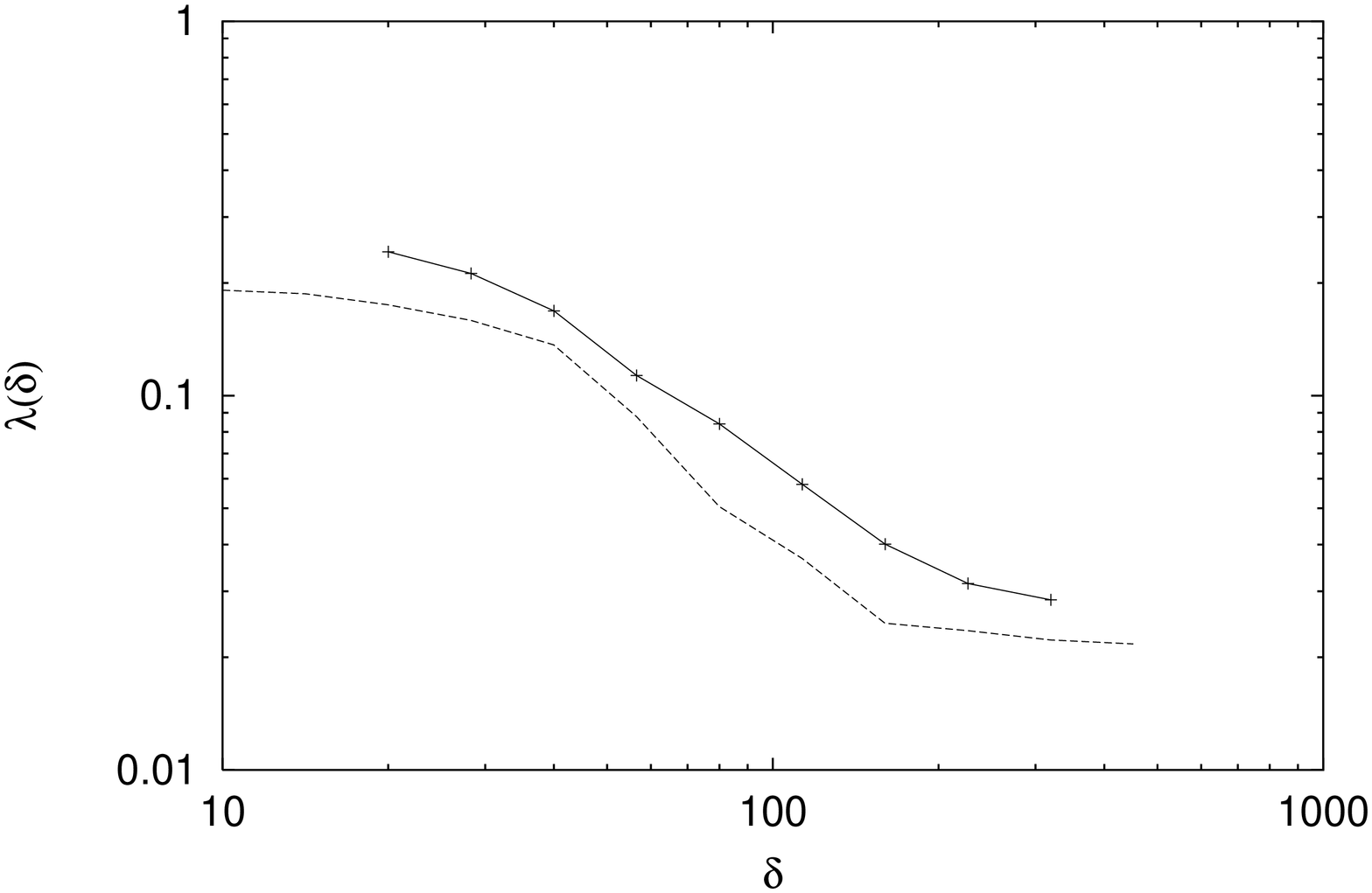,width=6cm,angle=270}}
\caption{}
\label{fig:6}
\end{figure}

\newpage

\begin{figure}[p]
\centerline{\epsfig{figure=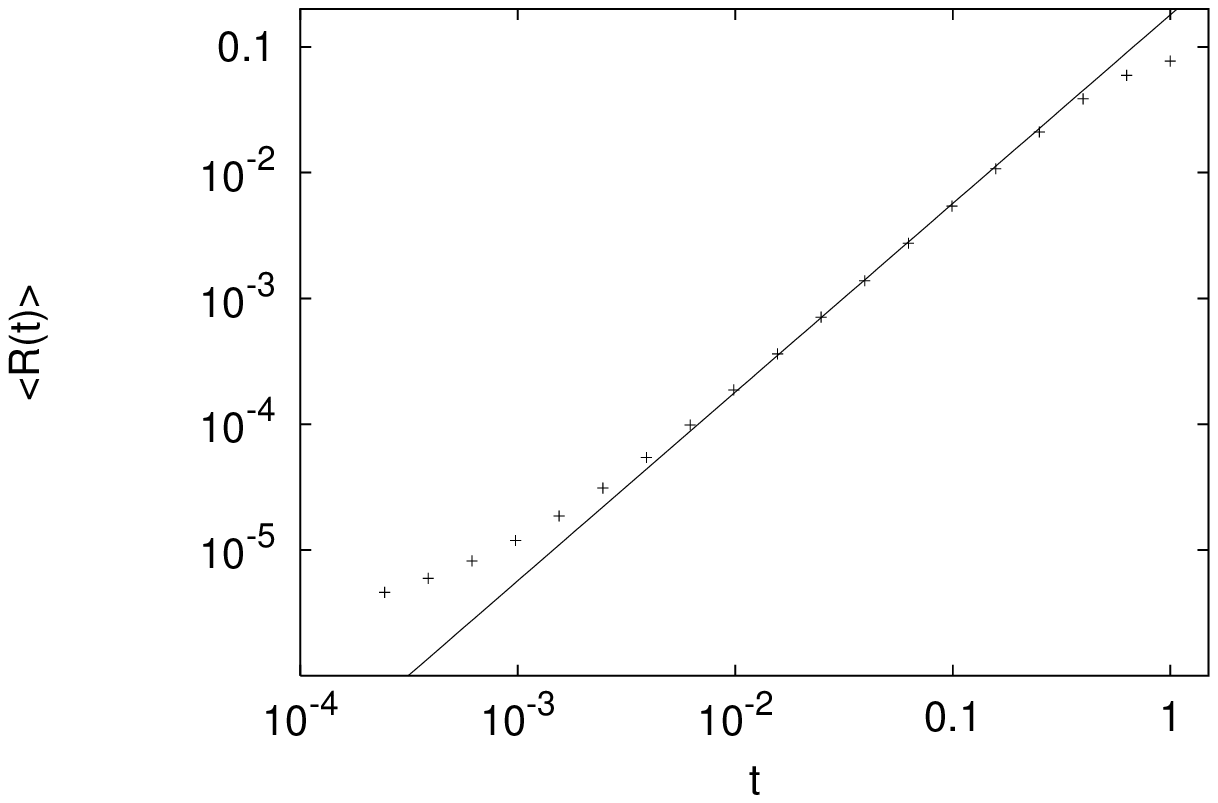,width=9cm,angle=0}}
\caption{}
\label{fig:7}
\end{figure}

\newpage

\begin{figure}[p]
\centerline{\epsfig{figure=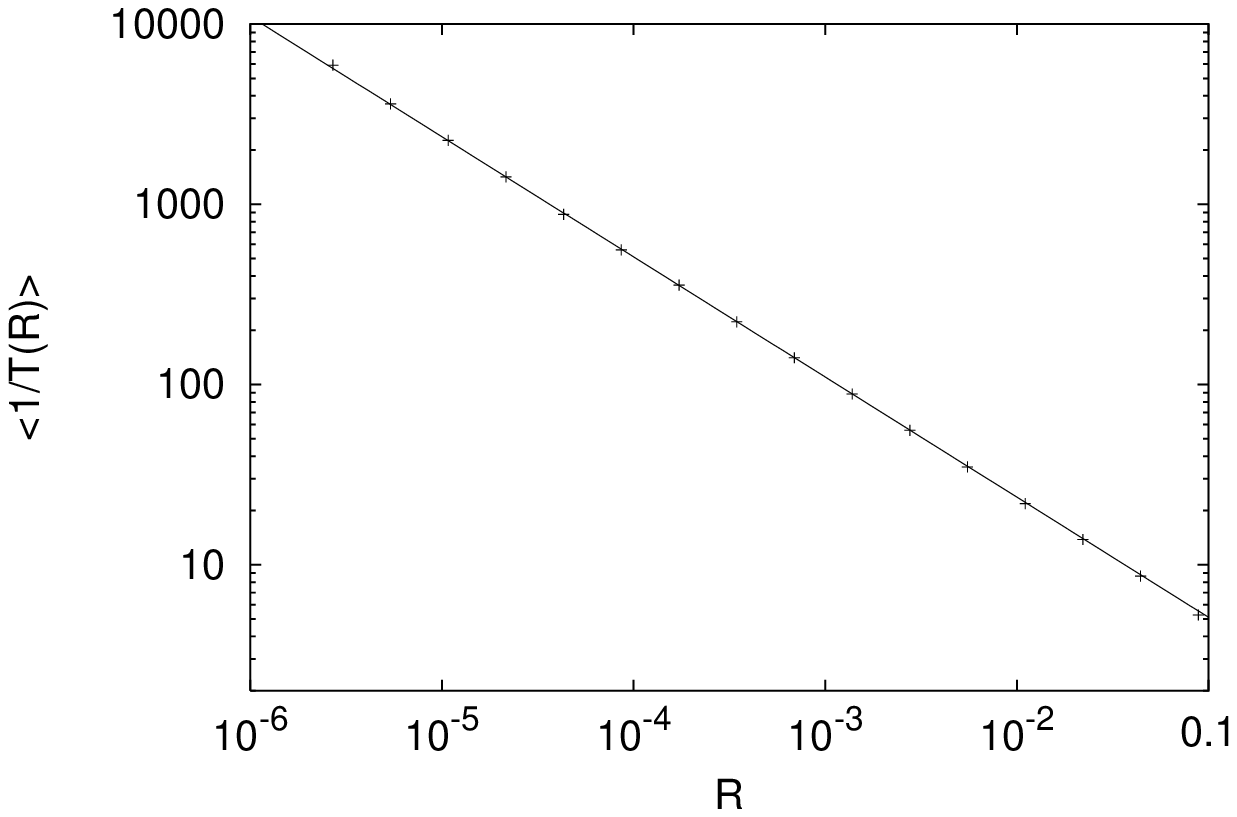,width=9cm,angle=0}}
\caption{}
\label{fig:8}
\end{figure}

\end{document}